\documentclass{article}
\usepackage[utf8]{inputenc}
\usepackage{graphicx}
\usepackage{amssymb}
\usepackage{url,lineno,microtype}
\usepackage{booktabs}
\usepackage{hyperref}
\usepackage{subcaption}
\usepackage{authblk}
\usepackage{accents}

\title{Circadian Rhythms are Not Captured Equal: Exploring Circadian Metrics Extracted by Different Computational Methods from Smartphone Accelerometer and GPS Sensors in Daily Life Tracking}
\author[1]{Congyu Wu\footnote{Correspondence: congyu.wu@austin.utexas.edu}}
\author[1]{Megan McMahon}
\author[2]{Hagen Fritz}
\author[1]{David M. Schnyer}
\affil[1]{Department of Psychology, University of Texas at Austin}
\affil[2]{Department of Civil, Environmental, and Architectural Engineering, University of Texas at Austin}

\date{July 2021}

\begin{document}

\maketitle

\begin{abstract}

Circadian rhythm is the natural biological cycle manifested in human daily routines. A regular and stable rhythm is found to be correlated with good physical and mental health. With the wide adoption of mobile and wearable technology, many types of sensor data, such as GPS and actigraphy, provide evidence for researchers to objectively quantify the circadian rhythm of a user and further use these quantified metrics of circadian rhythm to infer the user's health status. Researchers in computer science and psychology have investigated circadian rhythm using various mobile and wearable sensors in ecologically valid human sensing studies, but questions remain whether and how different data types produce different circadian rhythm results when simultaneously used to monitor a user. We hypothesize that different sensor data reveal different aspects of the user's daily behavior, thus producing different circadian rhythm patterns. In this paper we focus on two data types: GPS and accelerometer data from smartphones. We used smartphone data from 225 college student participants and applied four circadian rhythm characterization methods. We found significant and interesting discrepancies in the rhythmic patterns discovered among sensors, which suggests circadian rhythms discovered from different personal tracking sensors have different levels of sensitivity to device usage and aspects of daily behavior. 

\end{abstract}

\section{Introduction}\label{sec:intro}

Circadian Rhythm (CR) is defined as the periodic quality of an organism's activity that follows a 24-hour cycle. Medical research has found that healthy individuals' day-to-day life exhibits a strong circadian rhythm, whereas deviations from such a rhythm are often indicators of potential health issues \cite{karatsoreos2012effects}. Traditionally, circadian rhythm is assessed by questionnaires, such as the Morning-Eveningness Questionnaire (MEQ) \cite{horne1976self} and the Munich Chronotype Questionnaire (MCTQ) \cite{roenneberg2003life}. These questionnaires are centered around the respondent's own recollection of their bed and wake time. With increasing amounts and varieties of sensor data available from daily life tracking devices, we have unprecedented capability to evaluate circadian rhythm objectively and passively, requiring minimal efforts from the user. 

Multiple methods have been proposed for the purpose of CR characterization, using different sensor signals and different mathematical approaches. A straightforward approach is to automate the measurements of constructs solicited in the traditional CR surveys. As opposed to asking participants when a certain activity was performed in a day, we can instead find a digital marker in sensor signals that reflects the same event. For example, the first time in a day at which a person steps out of their home, while could be asked in a questionnaire, can be passively detected by their smartphone's GPS sensor. Higher on the scale of analytical complexity is the \textit{cosinor} method which was originally proposed in the 1960s \cite{halberg1969chronobiology} and continues to be improved upon \cite{marler2006sigmoidally}. Building on the assumption that a person's daily rest-activity pattern follows a sinusoidal shape, the cosinor method aims to fit a trigonometric function to the magnitude of a body-worn sensor signal (e.g., wrist-worn actigraphy) and extract coefficients that serve as CR descriptors such as amplitude and time of peak. To free CR analysis from complying to a rather strict sinusoidal assumption, non-parametric CR methods \cite{witting1990alterations, blume2016nparact} were proposed to directly compute metrics that quantify the variability of sensor signals within-day and across different days. Non-parametric methods have become popular due to their applicability to signals of all shapes without concerns of assumption violation. Last but not least, researchers have also employed Fourier analysis to characterize CR: by converting a signal to its frequency domain, one can examine the degree to which the signal follows a 24-hour cycle, thus quantifying the strength of its rhythmicity. 

The methods outlined above produce different metrics that describe related, but distinct aspects of the circadian pattern. While these methods have been individually proposed and implemented with the sensing technology and sensor data of choice by the researcher, we find that evidence is limited in the extant literature that elucidates (1) the inter-relations between the circadian metrics extracted by different methods and (2) the relationship between the same circadian metric extracted from different types of mobile sensing data collected by companionship technology such as smartphones and smartwatches.  

We aim to address these limitations in this paper. We base our analyses on smartphone tracking data we collected from 1584 college students in a major public university in the United States over three weeks in the 2018-2019 school year. We conduct CR analysis using the four major classes of methods outlined above (survey construct automation, cosinor, non-parametric, Fourier) on two types of smartphone sensor signals (GPS, accelerometer), and compare the same metrics extracted from different sensor signals as well as compare different metrics extracted from the same sensor signal. 

\section{Related Work}\label{sec:related}



As introduced in Section \ref{sec:intro}, the Morning-Eveningness Questionnaire (MEQ) \cite{horne1976self} and the Munich Chronotype Questionnaire (MCTQ) \cite{roenneberg2003life} are typically used to assess the circadian rhythm in participants' sleep and wake schedule. Another survey instrument, called the Social Rhythm Metric \cite{monk1990social}, quantifies the circadian rhythm in other activities such as social interaction, attending school/work, and having dinner. The commonality between these surveys is their focus on the time in a day when a participant completes a certain number or proportion of a certain type of activity. Schoedel et al. \cite{schoedel2020challenge} sought to passively detect participants' sleep and wake schedule using smartphone usage events such as the action of snoozing, alarm settings, and the first and last time of smartphone usage during a day. The authors also recorded timestamps at which the participants completed 25\%, 50\%, and 75\% of smartphone usage to track its progression throughout the day. We call this approach \textbf{survey construct automation}, because it directly operationalizes survey-solicited activities as sensor-detected events.

Trigonometric functions naturally allow the modeling of periodicity. Researchers have leveraged this to fit cosine curves to sensor signals' temporal distribution over a 24-hour period to characterize the circadian rhythm therein \cite{cho2019mood}. The method is interchangeably called trigonometric regression, harmonic regression, or \textbf{cosinor regression}, for which multiple days of continuously or equidistantly sampled sensor signals serve as the input data. In the most basic cosinor setup, signal value $X$ at time $t$ is modeled by function $X(t) = mesor + amplitude \cdot cos(2\pi[t - \phi]/24)$, in which three parameters of the fitted curve are of primary interest: (1) $mesor$: short for Midline Estimating Statistic Of Rhythm, representing the mean value of the curve across cycles of the circadian rhythm; (2) $amplitude$: one half of the range of the curve within a cycle, and (3) $\phi$ or \textit{acrophase}: the time in a cycle at which the curve reaches its maximum value. In a transformed version of the cosinor method \cite{marler2006sigmoidally}, researchers aim to better mimic signal distributions that have a flatter or sharper peak or valley area than a normal cosine curve, which is prevalent in human actigraphy data, by plugging the cosinor function as an independent variable into a sigmoid link function such as the logistic function. Transformed cosinor methods introduce additional coefficients through the link function but mesor, amplitude, and acrophase are still the key parameters describing the shape. Besides the model parameters, another important metric to pay attention to is the goodness-of-fit, often the F-statistic, which indicates the degree to which the cosinor model explains the variance in the signal over the cycles.

For sensor signals that are largely sinusoidal in shape, cosinor method has been the gold standard approach to characterize circadian rhythm. However, for many other signals that fail to satisfy such an assumption, a set of metrics extracted by \textbf{non-parametric methods} have become popular \cite{blume2016nparact}. The non-parametric methods also take as input equidistantly sampled (at least hourly) sensor signals but aim to directly quantify notions such as fragmentation within cycle, contrast between rest and activity within cycle, and stability between cycles. Table \ref{tab:nonpar} lists five typical non-parametric descriptors of circadian rhythm which can be computed for sensor signals of any shape. Of the five metrics, Intradaily Variability (IV) quantifies fragmentation within a daily cycle. Its value increases when there exists a higher amount of switching between low and high activity each day (``varies more"). Interdaily Stability (IS), on the other hand, characterizes the extent to which the signal pattern remains similar across different days and takes a greater value when such similarity is high. The other three metrics, M10, L5, and Relative Amplitude (RA) are based on the periods of time during a day when the signal sees the lowest and highest intensity. M10 is the mean signal value over the 10 consecutive hours during which such value is the highest, whereas L5 indicates the mean signal value over the 5 consecutive hours during which such value is the lowest. Relative Amplitude quantifies the contrast between M10 and L5: when L5 is very small relative to M10, the RA approaches 1 but becomes zero when the signal's temporal distribution is uniform.

\begin{table}[]
\renewcommand{\arraystretch}{1.5}
\begin{tabular}{@{ }lllll@{ }} 
\toprule
Descriptor & Formula & &  \\ \midrule
Intradaily Variability (IV) & $\frac{\Sigma_{i=2}^{n}(X_i-X_{i-1})^2/(n-1)}{\Sigma_{i=1}^{n}(X_i-\bar{X})^2/n}$ &  &  \\
Interdaily Stability (IS) &  $\frac{\Sigma_{h=1}^{24}(\bar{X}_h-\bar{X})^2/24}{\Sigma_{i=1}^n(X_i-\bar{X})^2/n}$ &  &  \\
Maximum 10-hour Activity (M10) & $\max_{h\in\{1,\ldots,24\}}[\Sigma_{h}^{min(h+9,24)}\bar{X}_h + \mathbf{1}(h>15)\Sigma_{1}^{h-15}\bar{X}_h ]/10$ &  &  \\ 
Minimum 5-hour Activity (L5) & $\min_{h\in\{1,\ldots,24\}}[\Sigma_{h}^{min(h+4,24)}\bar{X}_h +\mathbf{1}(h>20)\Sigma_{1}^{h-20}\bar{X}_h ]/5$ & & \\ 
Relative Amplitude (RA) & $\frac{M10-L5}{M10+L5}$ & & \\\bottomrule
\end{tabular}
\caption{Five non-parametric descriptors of circadian rhythm. In the formulas, $n$ represents the total number of sampled points. $X_i$ is the $i$-th sampled value of the signal. $\bar{X}$ is the mean value of all sampled points. $\bar{X}_h$ is the mean value of sampled points within hour $h \in \{1,\ldots,24\}$ across all days observed.} \label{tab:nonpar}
\end{table}

\textbf{Fourier analysis}, or spectral analysis, is another method to quantify circadian rhythm, specifically the strength of it. Using Fourier analysis, researchers first obtain the frequency spectrum of a signal, resulting in a periodogram. Then, the density of frequency that falls between a specific range can be used as an indicator of rhythmic strength, calculated as the area under the periodogram curve within the range. For circadian rhythm, this range should be a small neighborhood of 24 hours. Saeb et al. \cite{saeb2015mobile} first used this method on GPS location data collected from participants' smartphones. The authors created a Lomb-Scargle periodogram \cite{lomb1976least, scargle1982studies} for each participant's GPS coordinates, calculated the amount of energy that fell into the frequency bins within a 24±0.5-hour period, and used this measure as a circadian movement feature for further predictive modeling of depression symptoms. 

Other standalone studies exist that seek to quantify the notion of circadian pattern in unique ways. Abdullah et al. \cite{abdullah2016automatic} used smartphone features such as distance traveled to predict CR scores obtained from the Social Rhythm Metric survey. Canzian et al. \cite{canzian2015trajectories} proposed a ``routine index" which quantifies the similarity between the geographic distribution of an individual's smartphone-tracked location traces within a period of time during a day and that of the location traces within the same period of time during other days. Huckins et al. \cite{huckins2019fusing} also focused on the similarity concept and computed ``circadian similarity" scores which are intra-class correlations between day-to-day variations in smartphone sensing features extracted over the same daily epochs (e.g., morning, afternoon). As opposed to using a trigonometric function to approximate for rest and activity, Huang et al. \cite{huang2018hidden} used a Hidden Markov Model to infer latent activity states from participants' accelerometer signal and computed cosinor-like circadian metrics based on the curve generated by latent state probabilities.

\section{Research Questions}\label{sec:rq}

We find that different circadian rhythm metrics produced by different methods may be characteristics of similar aspects of circadian rhythm. The first aspect is the \textbf{temporal distribution} of a signal within cycle, describing the signal's behavior with respect to time within each day. Both the traditional CR surveys and the method of automating CR survey constructs using sensor data aim to describe the temporal distribution of a signal. The acrophase parameter ($\phi$) from cosinor regression and the non-parametric, Intradaily Variability measure also fall in this category. The second aspect of CR is the \textbf{activity span} of a signal within cycle, describing the signal's behavior with respect to magnitude. This aspect concerns all the amplitude related metrics including the mesor and amplitude parameters from the cosinor method as well as M10, L5, and Relative Amplitude from the non-parametric approach. Lastly, we find both the periodogram density (near 24 hours) from Fourier analysis and the non-parametric Interdaily Stability are characteristics of \textbf{circadian disruption} and both decrease when a signal deviates from a 24-hour cycle, which can manifest as significantly altered patterns on different days. 

\begin{itemize}
    \item \textbf{RQ1}: How does the intradaily temporal distribution of magnitude differ between smartphone accelerometer and GPS signals?
    \item \textbf{RQ2}: How does the level of circadian disruption differ between smartphone accelerometer and GPS signals?
    \item \textbf{RQ3}: How do different CR metrics correlate with one another in both smartphone accelerometer and GPS signals?
\end{itemize} 

We list our research questions above. Through answering RQ1 and RQ2, we examine the intradaily temporal distribution and the circadian disruption aspects of CR based on data from smartphone GPS and accelerometer, two heavily utilized sensors in ubiquitous computing research. We are interested in how CR manifests in the distinct yet related daily behavior captured by the two sensors: smartphone accelerometer data can reflect the user's interaction with their phone and physical activity (while the phone is carried on person), whereas GPS data reflects the user's location and mobility patterns. Magnitude can be straightforwardly calculated from accelerometer signals and is therefore used as their signal strength. GPS coordinates on the other hand do not straightforwardly lend to a signal magnitude measure, so we use \textit{displacement}, computed by differencing the GPS trace, as the signal strength of GPS. We will not compare GPS and accelerometer based on their activity span aspect of CR because of their inherently different definitions and scales of amplitude. Finally, through RQ3 we investigate the inter-relations between different CR metrics extracted from both smartphone accelerometer and GPS data.  

\section{Data}

The data we use for this study come from the UT1000 Project \cite{10.1093/gigascience/giab044} we conducted at the University of Texas at Austin in two deployments, one in the Fall 2018 semester and the other during Spring 2019 semester. In this project we collected smartphone tracking data from 1584 student participants (in total) for three weeks, which included GPS, accelerometer, and phone usage data from the participants' primary smartphones in addition to real-time survey data including participants' responses to daily activity, mood, and sleep questions. Specifically, the GPS data contain timestamped coordinates (longitude and latitude). The GPS sensor was configured to scan for one minute every 10-minute break. The accelerometer data are sampled at a frequency of 10 Hz and contain a reading along the X, Y, and Z axes relative to the smartphone's position at each sampling. Both GPS and accelerometer data are subject to hardware constraints such as phone power-off or user deactivation.

\section{Methods}\label{sec:method}

To create the required input for CR analysis, we performed data preprocessing as follows. For accelerometer data, we first calculate a magnitude value $\sqrt{X^2+Y^2+Z^2}$ for each accelerometer reading. Then we divide each hour into six 10-minute bins and calculate the mean absolute deviation from unit gravity (rest state accelerometer magnitude) for each bin to serve as the activity value. For a day with complete accelerometer data we get to create an accelerometer activity vector of $6*24=144$ magnitude deviation values. For GPS, we use the same 10-minute bins and calculate the mean coordinate within each bin. Then we compute the haversine distance in meters between the previous bin's mean coordinate and the current bin's mean coordinate and use the distance as the activity value of the current bin. Similarly, for a day with complete GPS data we get to create a GPS activity vector of $6*24=144$ displacement values. 

These 144-length vectors will be the input for further CR analysis. The CR methods outlined in Section \ref{sec:related} are highly sensitive to missing data and we do not want to fill in missing data by interpolation. As such, we retain participants who have at least five days of fully complete data (not necessarily consecutive), i.e., complete 144-length vectors of both accelerometer and GPS activity values. This is an extremely strict criterion because missing as little as 10 minutes worth of data on a day would disqualify the day as ``complete". Even so, our participant cohort is large enough that we can afford to restrict our dataset to the highest completeness of days and still provide a large enough sample size to conduct CR analysis. Our preprocessed dataset contains 162 participants from the 2018 deployment and 63 participants from 2019. 

For each of the days with complete accelerometer and GPS data from each of the participants retained, we perform the following operations for both accelerometer and GPS data. First, we compute the cumulative activity at the increment of 10 minutes and find the timestamps at which 25\%, 50\%, and 75\% of cumulative activity has been reached (``quartile activity timestamps"). Activity is defined as mean absolute deviation from gravity for accelerometer and displacement for GPS. Second, we fitted a transformed cosinor model with a logistic link function \cite{marler2006sigmoidally} to the signals to learn parameters such as mesor, amplitude, acrophase, and the F-statistic. Third, we implement the formulas of Interdaily Stability and Intradaily Variability (Table \ref{tab:nonpar}). Finally, we follow the circadian movement calculation by Saeb et al. \cite{saeb2015mobile} and computed the frequency energy that falls between 23.5 and 24.5 hours on the Lomb-Scargle periodogram. 

These computational efforts resulted in a list of metrics that we use to answer the research questions formulated in Section \ref{sec:rq}. To answer RQ1, we look into the differences between accelerometer and GPS in their quartile activity timestamps, acrophase and F-statistics from transformed cosinor modeling and Intradaily Variability. To answer RQ2, we look into the differences between accelerometer and GPS in their frequency energy and Interdaily Stability. For both RQ1 and RQ2, we use Welch's T test to compare values due to the likely different variance. 

To answer RQ3, we put together 11 CR metrics (three quartile activity times, amplitude/mesor/acrophase, IV/IS/RA, and periodogram power near a 24 hour cycle) and apply a mixed graphical model \cite{haslbeck2020mgm} to learn their inter-relations. Specifically, we fit a LASSO regression model for each of 11 CR metrics with the remaining 10 metrics as predictors, using a regularization parameter $\lambda$ optimized by 10-fold cross validation. Through this double-round-robin operation, the relation between each pair of outcomes receives a coefficient in two models. We take the average of the two coefficients and use that value as the inter-relation value between the corresponding pair of metrics. The reason we undergo this process rather than simply calculating a correlation matrix is that we want to use the regularization functionality of LASSO models to reduce the visibility of inconsequential correlations between the CR metrics and accentuate the prominent ones; due to the inherently related nature of the CR metrics, simply calculating the pairwise correlations will return many significant results. 

We repeat all procedures outlined above twice, once for the Fall 2018 group and once for the Spring 2019 group.

\begin{table}[]
\begin{tabular}{@{ }llll@{ }} 
\toprule
Fall 2018   & Accelerometer (deviation) & GPS (displacement) & p-value \\ \midrule
25\%-time   & 10.37 & 11.56 & \textbf{$<$0.001***} \\
50\%-time   & 14.56 & 14.87 & \textbf{0.018*} \\
75\%-time   & 18.51 & 17.89 & \textbf{$<$0.001***} \\
Acrophase   & 16.22 & 16.34 & 0.532 \\
F-statistic & 24.86 & 49.95 & \textbf{$<$0.001***} \\
IV          & 1.39 & 1.14 &  \textbf{$<$0.001***} \\
\bottomrule
\end{tabular}
\caption{Comparing CR metrics that describe the intradaily temporal distribution of smartphone accelerometer and GPS activity, Fall 2018 participants.} \label{tab:result_temp_2018}
\bigskip
\begin{tabular}{@{ }llll@{ }} 
\toprule
Spring 2019 & Accelerometer (deviation) & GPS (displacement) & p-value \\ \midrule
25\%-time   & 10.45 & 11.2 & \textbf{0.011*} \\
50\%-time   & 14.89 & 14.63 & 0.235 \\
75\%-time   & 18.65 & 17.77 & \textbf{$<$0.001***} \\
Acrophase   & 16.7 & 16.25 & 0.187 \\
F-statistic & 44.84 & 102.86 & \textbf{$<$0.001***} \\
IV          & 1.36 & 0.92 &  \textbf{$<$0.001***} \\
\bottomrule
\end{tabular}
\caption{Comparing CR metrics that describe the intradaily temporal distribution of smartphone accelerometer and GPS activity, Spring 2019 participants.} \label{tab:result_temp_2019}
\end{table}

\section{Results}

\subsection{RQ1} 

Tables \ref{tab:result_temp_2018} and \ref{tab:result_temp_2019} show the results of comparing CR metrics that describe the intradaily temporal distribution of smartphone accelerometer and GPS activity, from the Fall 2018 and the Spring 2019 participants respectively. In both participant groups, the 25\% activity time is significantly earlier in accelerometer than GPS: the cumulative accelerometer activity reaches the first quartile between 10-11am while the cumulative GPS activity reaches the first quartile between 11am and noon. The 50\% or median activity time in accelerometer and GPS are quite similar in both groups, both falling between 2-3pm. Even though in the Fall 2018 group, the median activity time in GPS is significantly later than accelerometer (14.87 $>$ 14.56, p=0.018), such significance is not reproduced in the 2019 group. The 75\% activity time registers a significant difference between accelerometer and GPS again, with the latter significantly earlier. In both groups, cumulative accelerometer activity reaches 75\% of the day between 6-7pm while GPS activity does so between 5-6pm. 

Moving on to the cosinor metrics, we see that the F-statistic is significantly greater for GPS than accelerometer in both groups, indicating that daily GPS activity follows a sinusoidal shape more closely than accelerometer. However, the acrophase, or predicted time of peak activity, shows little and insignificant difference between accelerometer and GPS in both groups. This indicates that the time when mobility behavior peaks and the time when physical activity peaks during a day converge. Note the clear difference between the acrophase time and the 50\% activity time in both groups and both sensors: acrophase falls between 4-5pm while 50\% activity time falls between 2-3pm. This temporal shift between median cumulative activity and peak activity during a day is tantamount to the difference between median and mode in the statistical context. The later acrophase than the 50\% activity time suggests a left skewed daily activity profile in both accelerometer and GPS.  

Last but not least, Intradaily Variability is significantly greater in accelerometer than GPS in both groups, indicating that there exists more severe fragmentation, or switching between rest and activity, in accelerometer than GPS. This finding is not surprising because an individual tends to interact with their smartphone in ``fragments" -- picking it up, using it for a while, then leaving it unattended until the next use -- thus contributing to bouts of activity between periods of rest. However, in one's mobility pattern, such fragmentation is subject to the actual number of places the person visits and stays at during the day, which requires more effort and thus less likely to happen frequently. 

\begin{table}[]
\begin{tabular}{@{ }llll@{ }} 
\toprule
Fall 2018        & Accelerometer (deviation) & GPS (displacement) & p-value \\ \midrule
24±0.5h energy   & 26.42 & 18.04 & \textbf{$<$0.001***} \\
IS              & 0.26 & 0.09 & \textbf{$<$0.001***} \\
\bottomrule
\end{tabular}
\caption{Comparing CR metrics that describe the level of circadian disruption in smartphone accelerometer and GPS activity, Fall 2018 participants.} \label{tab:result_disr_2018}
\bigskip
\begin{tabular}{@{ }llll@{ }} 
\toprule
Spring 2019      & Accelerometer (deviation) & GPS (displacement) & p-value \\ \midrule
24±0.5h energy   &  32.60 & 22.31 & \textbf{$<$0.001***} \\
IS               &  0.22 & 0.04 & \textbf{$<$0.001***} \\
\bottomrule
\end{tabular}
\caption{Comparing CR metrics that describe the level of circadian disruption in smartphone accelerometer and GPS activity, Spring 2019 participants.} \label{tab:result_disr_2019}
\end{table}

\subsection{RQ2} 

Tables \ref{tab:result_disr_2018} and \ref{tab:result_disr_2019} show the results of comparing the level of circadian disruption in smartphone accelerometer and GPS activity, from the Fall 2018 and the Spring 2019 participants respectively. Both the energy within 24±0.5 hours on the Lomb-Scargle periodogram and Interdaily Stability are significantly higher in accelerometer than GPS for both participant groups. This result provides a straightforward answer to RQ2: there is a significantly stronger circadian rhythm, or a lower level of circadian disruption in daily accelerometer activity than GPS. The difference between Interdaily Stability between accelerometer and GPS is direct evidence that our participants followed a more homogeneous schedule of phone use and/or physical activity on different days as opposed to mobility patterns, which could vary greatly on different days depending the daily agenda.

\begin{figure}
     \centering
     \begin{subfigure}[b]{0.48\textwidth}
         \centering
         \includegraphics[width=\textwidth]{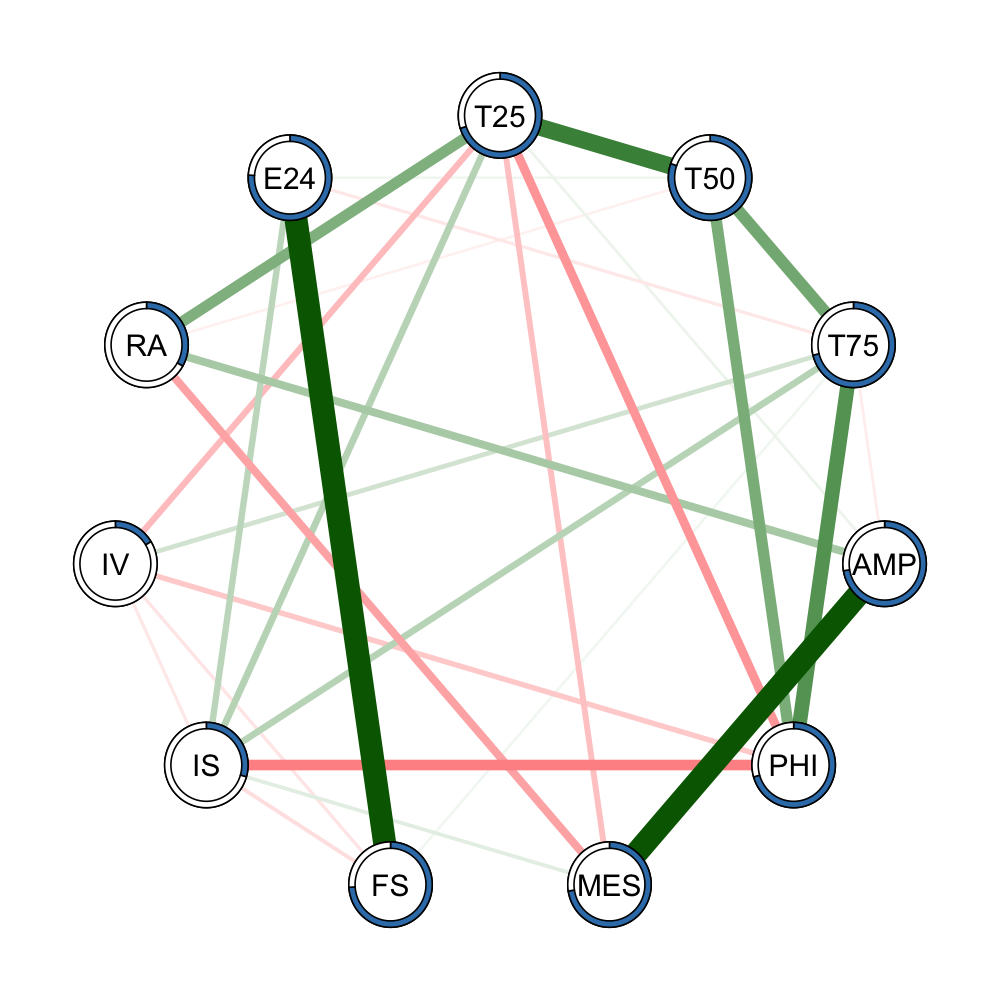}
         \caption{Accelerometer, Fall 2018}
         \label{fig:acc_2018}
     \end{subfigure}
     \begin{subfigure}[b]{0.48\textwidth}
         \centering
         \includegraphics[width=\textwidth]{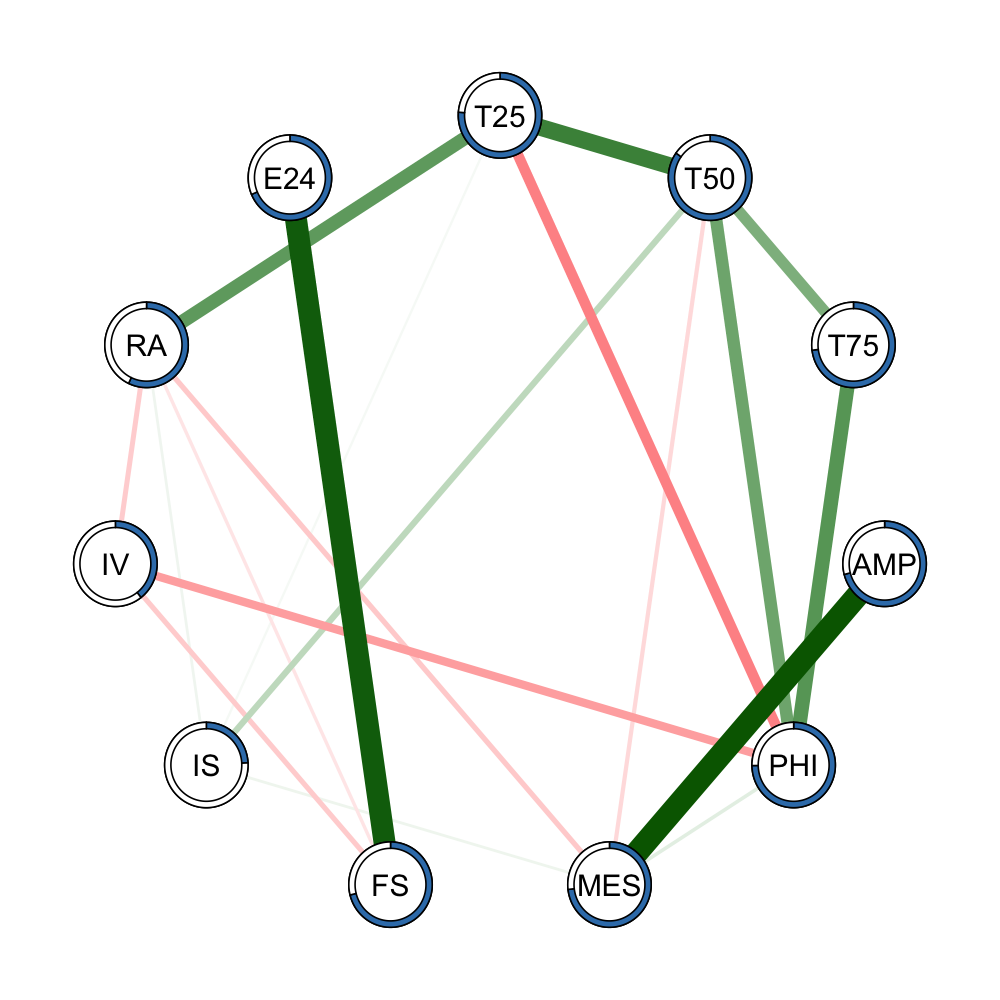}
         \caption{Accelerometer, Spring 2019}
         \label{fig:acc_2019}
     \end{subfigure}
     \hfill
     \begin{subfigure}[b]{0.48\textwidth}
         \centering
         \includegraphics[width=\textwidth]{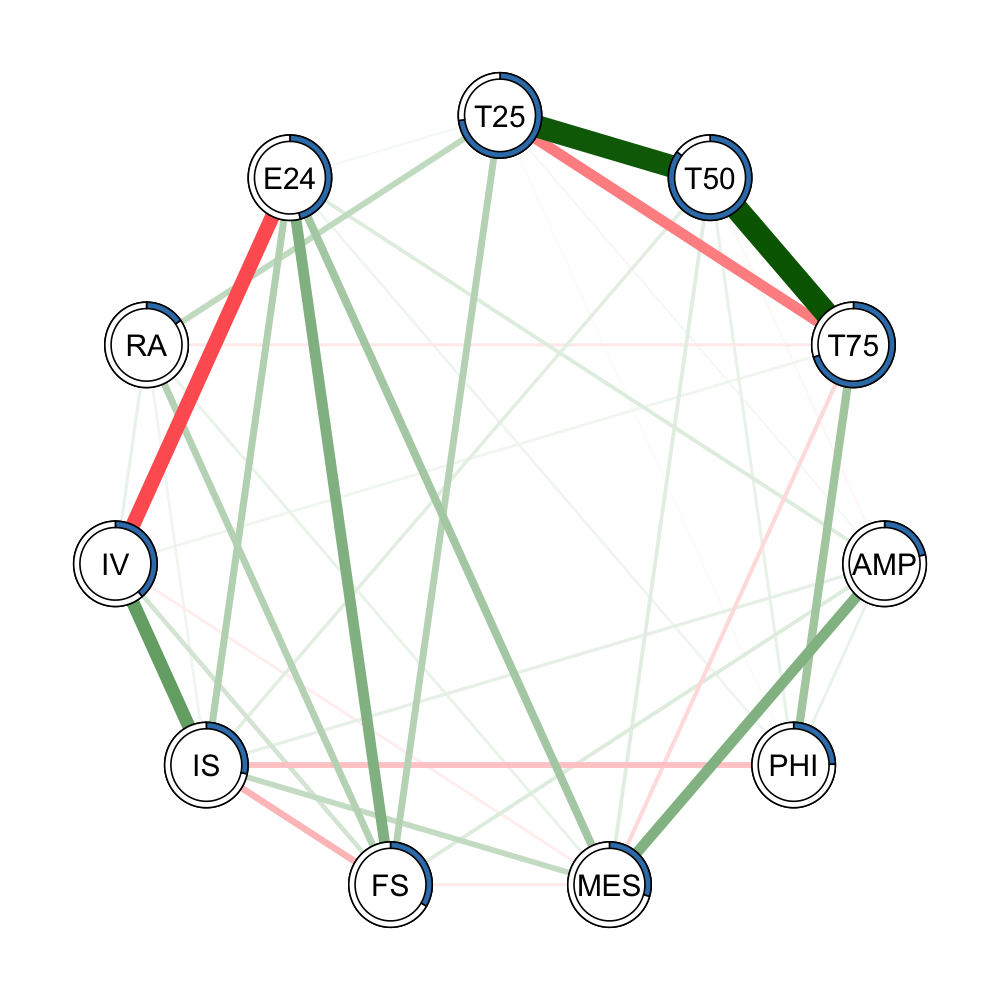}
         \caption{GPS, Fall 2018}
         \label{fig:gps_2018}
     \end{subfigure}
     \begin{subfigure}[b]{0.48\textwidth}
         \centering
         \includegraphics[width=\textwidth]{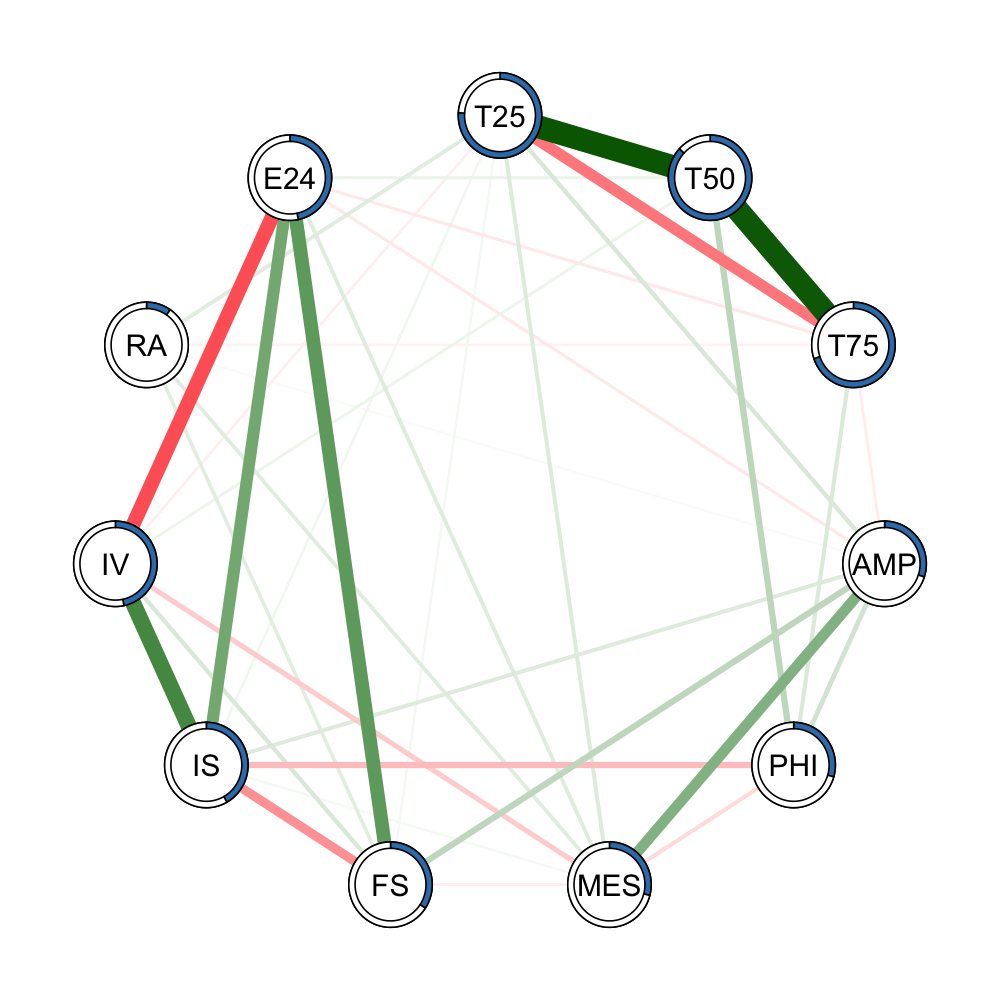}
         \caption{GPS, Spring 2019}
         \label{fig:gps_2019}
     \end{subfigure}
        \caption{Inter-relations of CR metrics from smartphone accelerometer and GPS activity (see Section \ref{sec:method} for details). Each node represents a CR metric and edges represent relations between metrics. Color and weight of the edges indicate the direction (green: positive; red: negative) and magnitude of the pairwise relation. The blue part in the ring surrounding each node indicates the proportion of variance explained by other metrics. T25/T50/T75: time of 25\%/50\%/75\% cumulative activity; AMP/PHI/MES/FS: amplitude, acrophase, mesor, and F-statistics from the transformed cosinor modeling; IV/IS/RA: Intradaily Variability, Interdaily Stability, and Relative Amplitude from the non-parametric methods; E24: energy of frequency that falls within 24±0.5 hours on the Lomb-Scargle periodogram.}
        \label{fig:RQ3}
\end{figure}

\subsection{RQ3}

Figure \ref{fig:RQ3} shows the inter-relations between 11 CR metrics extracted by four types of methods from smartphone accelerometer and GPS data collected from both the 2018 and the 2019 participant groups. Each node represents a CR metric and the edges in-between represent their inter-relations. The edges' color indicates the sign of the pairwise relation (green positive, red negative) and thickness is proportional to the magnitude. Overall, the inter-relations between CR metrics in both accelerometer and GPS are well-replicated between the two participant groups, with the 2019 group seeing slightly fewer inter-relations between the CR metrics in general. 

Among the three quartile activity times, 50\% activity time is to a large extent explained by 25\% and 75\% activity times in both accelerometer and GPS, meaning that a person who arrives at median cumulative activity later (or earlier) in the day tends to complete 25\% and 75\% of sensor activity later (or earlier) as well. A surprising negative link exists between 25\% time and 75\% time unique to GPS, suggesting that in terms of mobility activity the participants who started their days early tend to be the ones who finished their days late.

Among the cosinor modeling measures, the only relation that stands out is a strongly positive one between mesor (MES) and amplitude (AMP). This is an expected pattern since both mesor and amplitude describe the signal's activity span and are supposed to be codependent. Interestingly, neither acrophase (PHI) nor the F-statistic (FS) is related to mesor or amplitude. This means that the time when accelerometer and GPS activity reach a peak and how closely the cosinor model fits activity data are both independent of how large the peak is.

The inter-relations between the three non-parametric CR metrics (IV, IS, RA) seem very weak in both accelerometer and GPS, except for a strong, positive association between IV and IS evident only in GPS. Also unique to GPS is a strong negative relation between IV and E24, the energy of frequency that falls between 23.5 and 24.5 hours on the Lomb-Scargle periodogram, of which we see none in accelerometer. This indicates that participants who have higher Intradaily Variability in mobility tend to have more disrupted CR in mobility across days. 

Lastly, there exists a consistent positive relation between E24 and the F-statistics from cosinor modeling in both accelerometer and GPS, with that in accelerometer even stronger. This relation is evidence that better fit with the cosinor model translates to stronger rhythmicity with a 24-hour cycle.

\section{Discussion}

The relationship between the daily behavioral aspects captured by smartphone accelerometer and GPS is worth noting. Straightforwardly, the GPS sensor detects changes in geographic location, therefore captures the user's place visits and mobility patterns. However, we recognize at least three types of behavior that can lead to increased accelerometer signal activity (or greater deviation from gravity baseline): (1) interacting with the phone such as making phone calls and using mobile apps; (2) performing physical activities with phone carried on person while staying at the same place, such as cooking at home or walking around in an office, and; (3) traveling between places with phone carried on person, which is also reflected in GPS signals. In this sense, accelerometer captures a wider range of daily behavior than GPS. 

\section{Conclusion}

In this study, we used smartphone tracking data from 225 college student participants with at least five full days of complete data to investigate inter-relations between circadian patterns extracted from accelerometer and GPS sensors. We used four analytical approaches, namely survey construct automation, transformed cosinor modeling, non-parametric methods, and Fourier analysis, to characterize circadian rhythm using both data types. We conceptually categorized CR metrics into three categories, namely temporal distribution, activity span, and circadian disruption. We asked three research questions: RQ1 and RQ2 compared CR metrics of the temporal distribution category and the circadian disruption category between smartphone accelerometer and GPS activity, whereas RQ3 examined the inter-relations between different CR metrics within the same sensor.

We found that, compared to GPS signals, the intradaily distribution of smartphone accelerometer activity follows a pattern that starts earlier in the day; winds down later; reaches half cumulative activity about the same time, which is about two hours earlier than when it reaches maximum activity; conforms less to a sinusoidal wave; and exhibits more intradaily fragmentation. Moreover, GPS activity exhibited a stronger circadian rhythm and interdaily stability than accelerometer, revealing differences in the daily behavioral aspects that manifest in the two sensor signals. Finally, the inter-relations between different CR metrics have their respective peculiarities in accelerometer and GPS activity. A notable one is the negative relation between intradaily variability and circadian rhythm strength, which is present in GPS but not in accelerometer. 

Our explorations reported in this paper demonstrated the reactivity of passive circadian monitoring and computation to mobile sensor choices (thus daily behavioral aspects) and offered empirical evidence for the inter-relations between a comprehensive set of different circadian rhythm measures.

\section*{Funding}
This work was supported by Whole Communities—Whole Health, a research grand challenge at the University of Texas at Austin.

\bibliographystyle{plain}
\bibliography{main}

\end{document}